# Optomechanical dark-mode-breaking cooling

Yan Cao[1,†], Cheng Yang[1,†], Jiteng Sheng[1,2,*], and Haibin Wu[1,2,3,*]

[1]State Key Laboratory of Precision Spectroscopy, Institute of Quantum Science and Precision Measurement, East China Normal University, Shanghai 200062, China

[2]Collaborative Innovation Center of Extreme Optics, Shanxi University, Taiyuan 030006, China

[3]Shanghai Research Center for Quantum Sciences, Shanghai 201315, China

†*These authors contributed equally to this work.*

*jtsheng@lps.ecnu.edu.cn; hbwu@phy.ecnu.edu.cn*

**Optomechanical cooling of multiple degenerate mechanical modes is prevented by the mechanical dark mode due to destructive interference. Here we report the first experimental demonstration of simultaneous cooling of two near-degenerate mechanical modes by breaking the mechanical dark mode in a two-membrane cavity optomechanical system. The dark mode is generated as the system passes the exceptional point of the anti-parity-time symmetric scheme. By introducing a second cavity mode for the additional dissipative channel, the dark mode is broken and the total phonon number is reduced by more than an order of magnitude below the dark mode cooling limit. Owing to the flexible tunability of the optomechanical coupling rates of such a four-mode coupled system, the optimized cooling efficiency can be achieved by investigating different parameter ranges. Our results provide an important step towards the ground state cooling and entanglement among multiple degenerate mechanical resonators.**

Cavity optomechanics provides powerful platforms for studying macroscopic quantum physics by utilizing mutual interaction between electromagnetic radiation and mechanical motion [1]. Remarkable accomplishments have been achieved with single-mode cavity optomechanical systems, i.e., one cavity mode interacts with one mechanical mode, including such as ground state cooling [2,3], optomechanical squeezing [4-6], and quantum interface [7-9]. Recently, cavity optomechanical systems with multiple mechanical modes have received much attention, since they can investigate important physics that are not available in single-mode systems, for example, macroscopic entanglement [10,11]. Although optomechanical cooling and entanglement of multiple mechanical resonators can be achieved by selecting the mechanical frequency difference to be larger than the optomechanical coupling rate [10-12], ground state cooling and entanglement of multiple degenerate mechanical resonators are crucial in many studies, e.g. quantum gravity tabletop experiments [13-16] and entanglement enhancement [17,18]. However, the dark mode is the obstacle that has to be overcome in order to achieve the simultaneous cooling of multiple degenerate or near-degenerate mechanical modes [19-21]. Dark mode or dark state, which typically refers to a system that is prepared in a particular state and decouples from its surroundings, is acknowledged as a universal phenomenon that exists in a large variety of systems [22-28]. Although a great deal of effort has been devoted to the proposals for the optomechanical cooling of multiple degenerate mechanical resonators, e.g. multiple dissipation channels [29,30], nonreciprocal energy transfer [31], reservoir engineering [32], feedback cooling [33], Lyapunov control [34], and cross-Kerr nonlinearity [35], the experimental demonstration remains elusive.

In this work, we report an experimental demonstration of optomechanical cooling of two near-degenerate mechanical modes by breaking the dark mode in a two-membrane cavity optomechanical system. In contrast to a typical two-membrane cavity optomechanical system [36-39], where two nanomechanical membranes interact with a common cavity field, an additional distinguishable cavity mode is introduced as an auxiliary dissipative channel [29,30]. We have directly observed the processes of the mechanical dark mode generation and breaking. Thus, the simultaneous cooling is realized beyond the dark mode cooling limit. In addition, the cooling efficiency can be optimized by utilizing the advantages of the flexible tunability of the system parameters. Our results provide a foundation for achieving ground state cooling and macroscopic quantum entanglement of degenerate systems, and also can be straightforwardly extended to higher-dimensional optomechanical systems.

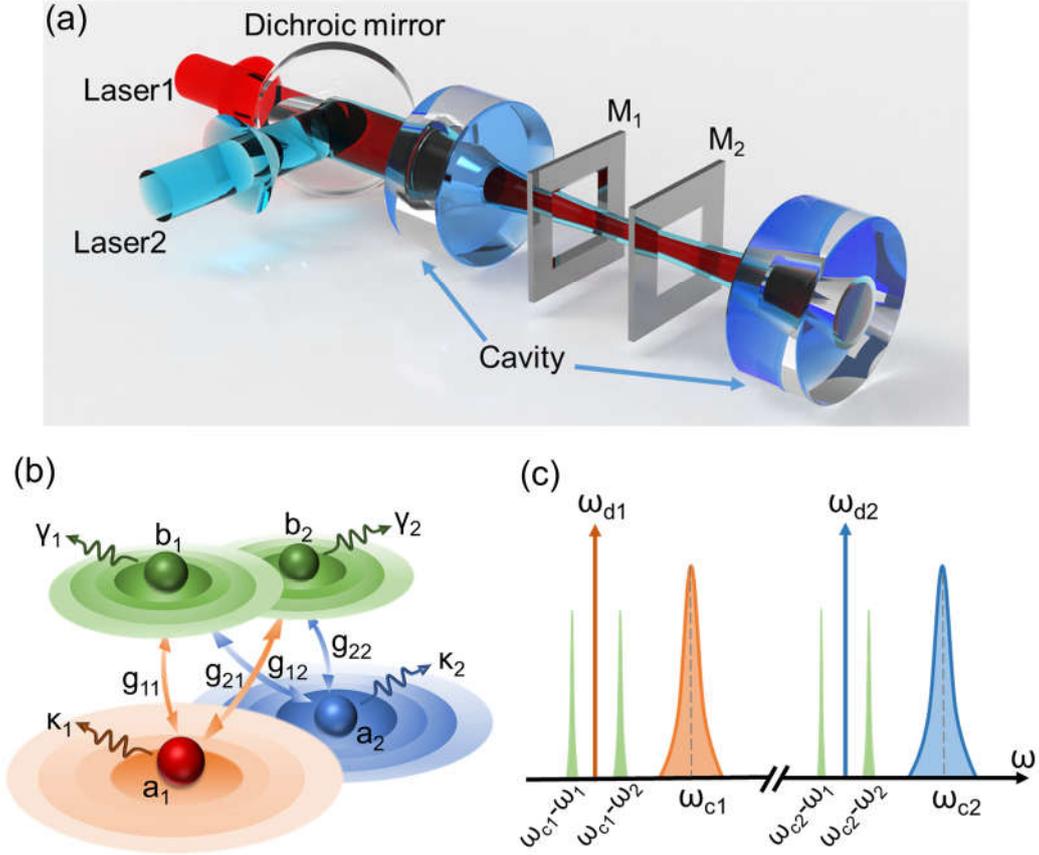

**Fig. 1.** (a) Schematic illustration of the two-membrane cavity optomechanical system for the optomechanical dark-mode-breaking cooling. Two nanomechanical membranes ($M_1$ and $M_2$) are placed inside an optical cavity with two distinct cavity modes driven by two different color lasers. (b) Schematic diagram of the multimode optomechanical system formed by two cavity modes $a_1$ and $a_2$ (with optical amplitude decay rates $\kappa_1$ and $\kappa_2$) optomechanically coupled to two mechanical modes $b_1$ and $b_2$ (with mechanical amplitude decay rates $\gamma_1$ and $\gamma_2$), respectively. $g_{ij}$ (i, j = 1,2) represents the single-photon optomechanical coupling strength between i-th mechanical mode and j-th cavity mode. (c) Frequency domain of cooling two near-degenerate mechanical modes with two optical modes. The optomechanical system consists of two mechanical modes with frequencies $\omega_1$ and $\omega_2$, and two cavity modes with $\omega_{c1}$ and $\omega_{c2}$. The optomechanical system is driven by two distinct optical fields at $\omega_{d1}$ and $\omega_{d2}$.

Our experiment is performed in a two-membrane cavity optomechanical system driven by two distinct optical fields, as shown in Fig. 1a [40]. Two spatially separated silicon nitride membranes are placed inside a high-finesse Fabry-Perot cavity with a free spectral range of ~ 4.2 GHz. The linewidths of the optical cavity with two membranes in the middle are about 270 kHz for laser 1 (with a wavelength of 1064 nm) and 290 kHz for laser 2 (with a wavelength of 795 nm). We utilize the vibrational (3,3) modes of the membranes, which are nearly degenerate with frequencies $\omega_1 \sim \omega_2 \sim 2\pi \times 1.2$ MHz, and have the mechanical amplitude decay rates $\gamma_1 = 2\pi \times 0.65$ Hz and $\gamma_2 = 2\pi \times 0.62$ Hz. The optomechanical system is in the sideband-resolved regime. The mechanical frequency difference, $\delta\omega = \omega_1 - \omega_2$, can be tuned to be degenerate or near-degenerate via the piezos where the membranes are attached [36]. Laser 1 and laser 2 drive two widely separated cavity modes (at frequencies $\omega_{c1}$ and $\omega_{c2}$) in the red-detuned regime, with the driven frequencies $\omega_{d1,d2} = \omega_{c1,c2} - (\omega_1 + \omega_2)/2$, as shown in Fig. 1c. $g_{ij}$ represents the single-photon optomechanical coupling strength between i-th mechanical mode and j-th cavity mode (see Fig. 1b), which is tunable in the range of $\pm 2\pi \times 1$ Hz by controlling the relative position of the corresponding membrane along the cavity axis. The mean photon numbers inside the optical modes could reach up to $10^9$ at the input laser power of about 10 mW. The motions of two mechanical resonators are measured using the optical beam deflection method with a single probe beam from an additional 1550 nm laser that passes through the cavity (not shown in Fig. 1a) due to low cavity mirror reflectivity at this wavelength. This allows for the extraction of the linear combination of two mechanical resonators' displacements in comparison to the use of two independent beams for measuring individual membrane motions [37].

Following the standard procedure, the Hamiltonian of a cavity optomechanical system can be linearized by splitting the optical mode $a_j$ into the average coherent amplitude $\alpha_j$ and the fluctuating term $\delta a_j$, i.e. $a_j = \alpha_j + \delta a_j$ [1] in the red-detuned regime. Therefore, the linearized Hamiltonian of the optomechanical system with two mechanical modes $b_i$ and two cavity modes $a_j$ under the rotating wave approximation is given with a beam splitter interaction by [40]

$$H = \sum_{i=1}^{2} \omega_i b_i^\dagger b_i - \sum_{j=1}^{2} \Delta_j \delta a_j^\dagger \delta a_j - \sum_{i=1}^{2} \sum_{j=1}^{2} G_{ij} \left( \delta a_j^\dagger b_i + \delta a_j b_i^\dagger \right). \quad (1)$$

Here $G_{ij} = g_{ij}\alpha_j$ is the multiphoton optomechanical coupling strength which is assumed to be real [1]. $\Delta_j = \omega_{dj} - \omega_{cj}$ is the frequency detuning between the driving laser and the cavity field. The mechanical dark mode can be generated when two mechanical modes are coupled to a single cavity mode, e.g. $G_{11} \neq 0$, $G_{21} \neq 0$, and $G_{12} = G_{22} = 0$. In this case, we can define a mechanical bright mode $b_+ = (G_{11}b_1 + G_{21}b_2)/(G_{11}^2 + G_{21}^2)^{1/2}$ and a mechanical dark mode $b_- = (G_{21}b_1 - G_{11}b_2)/(G_{11}^2 + G_{21}^2)^{1/2}$ [31]. When the mechanical frequencies are near-degenerate and the optomechanical coupling strength is strong enough, i.e., $|G_{11}G_{21}|/\kappa_1 \gg |\delta\omega|/2$, the coupling between the mechanical dark mode (decoupled from the optical field) and bright mode can be neglected, preventing the further optomechanical cooling [21]. The dark mode in an optomechanical system with one cavity mode and two mechanical modes can be analogous to the dark state in a three-level $\Lambda$-type atomic system, which has been utilized for optomechanical stimulated Raman adiabatic passage [46].

By eliminating the cavity modes adiabatically, the effective cavity-mediated dynamics of two mechanical resonators is governed by the effective Hamiltonian [47]

$$H_{eff} = \begin{pmatrix} \omega_1 - i\gamma_1 - i\Gamma_{11} - i\Gamma_{12} & -i\Gamma_1 - i\Gamma_2 \\ -i\Gamma_1 - i\Gamma_2 & \omega_2 - i\gamma_2 - i\Gamma_{21} - i\Gamma_{22} \end{pmatrix}. \quad (2)$$

The diagonal elements consist of $\Gamma_{11} = -g_{11}^2 \text{Im}(\chi_1^{eff})$, $\Gamma_{12} = -g_{12}^2 \text{Im}(\chi_2^{eff})$, $\Gamma_{21} = -g_{21}^2 \text{Im}(\chi_1^{eff})$, and $\Gamma_{22} = -g_{22}^2 \text{Im}(\chi_2^{eff})$, which represent the dynamical backaction of two cavity modes on each mechanical resonator. The non-diagonal elements $\Gamma_1 = -g_{11}g_{21}\text{Im}(\chi_1^{eff})$ and $\Gamma_2 = -g_{12}g_{22}\text{Im}(\chi_2^{eff})$ are the cavity-mediated coupling between two mechanical modes. $\chi_{1,2}^{eff}$ is the effective susceptibility introduced by the corresponding cavity mode, which is dominantly imaginary, and can be controlled by the cavity input power $P_{1,2}^{in}$. Here the optical spring terms corresponding to the real part of the effective susceptibilities $\chi_{1,2}^{eff}$ have been omitted when $|\Delta_j| \approx \omega_i (i, j = 1, 2)$ in the resolved sideband regime [40].

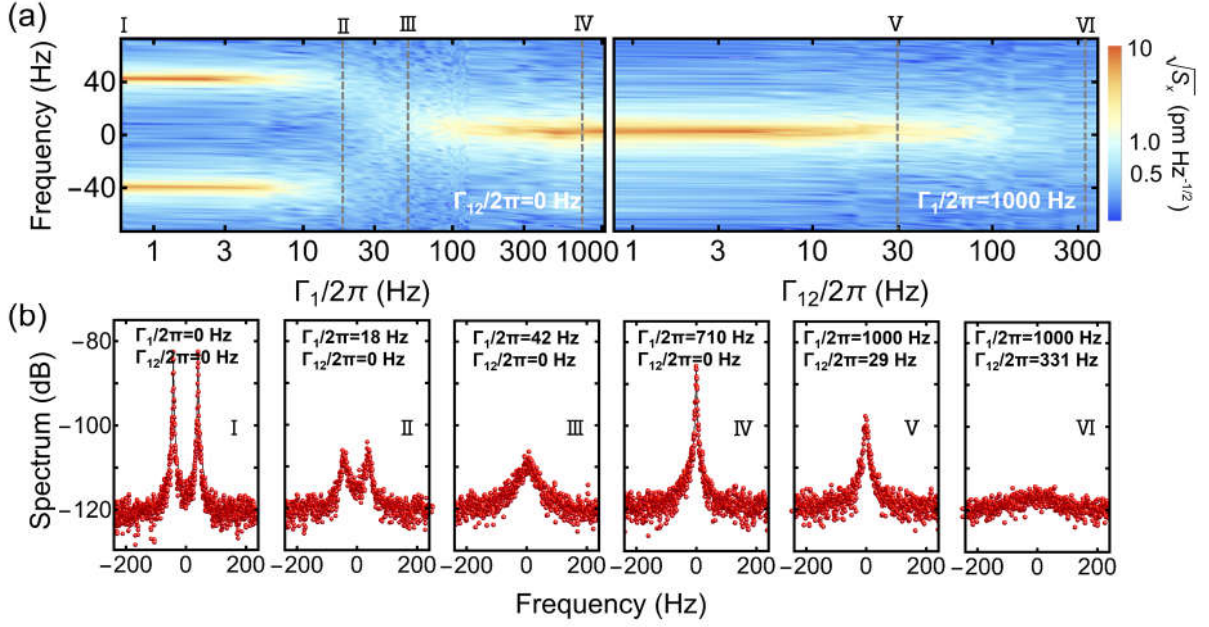

**Fig. 2.** (a) Measured mechanical power spectral density (PSD) for the dark mode generation (left panel) and the dark mode breaking (right panel) detected by the 1550 nm probe beam. The colorbar represents the square root of PSD $S_x$. (b) The corresponding mechanical spectra at different $\Gamma_1$ and $\Gamma_{12}$, marked as I-VI in Fig. 2a.

In order to clearly observe the processes of the dark mode generation and dark mode breaking, we first study a special situation that $G_{11} \neq 0$, $G_{21} \neq 0$, $G_{12} \neq 0$, and $G_{22} = 0$, by precisely controlling the single-photon optomechanical coupling strengths to satisfy the conditions $g_{11} = g_{21} \neq 0$, $g_{12} \neq 0$, and $g_{22} = 0$. Therefore, the cavity mode $a_2$ provides an additional nonzero dissipation term $\Gamma_{12}$ for the mechanical mode $b_1$. Moreover, the mechanical frequency difference is tuned to be $\delta\omega = 2\pi \times 80$ Hz. Figure 2a depicts the power spectral density (PSD) of mechanical resonators for the dark mode generation (left panel) and the dark mode breaking (right panel), and Fig. 2b shows the corresponding mechanical spectra at different $\Gamma_1$ and $\Gamma_{12}$. As illustrated in the left panel of Fig. 2a, the amplitudes of two mechanical modes decrease and the linewidths become larger as $\Gamma_1$ increases due to the optomechanical cooling effect. The mechanical modes attract and coalesce when the system exceeds the exceptional point (EP) of such an anti-parity-time symmetric scheme [39,48] at $\Gamma_1/2\pi = 40$ Hz. By further increasing $\Gamma_1$, the amplitude of the mechanical mode becomes larger instead of smaller (see III and IV in Fig. 2b), indicating the generation of the mechanical dark mode. In order to break the dark mode, we introduce another cavity mode $a_2$ as the additional dissipative channel. By keeping $\Gamma_1/2\pi$ = 1000 Hz and increasing $\Gamma_{12}$ from zero, the amplitude of the mechanical mode decreases, as shown in the right panel of Fig. 2b, which behaves similarly to typical resolved sideband cooling of a single mechanical resonator [1].

By fitting the mechanical spectra in Fig. 2 with the standard Lorentzian function, both the frequency and linewidth of the eigenmodes can be extracted, which are plotted in Figs. 3a and 3b, respectively. At a relatively small $\Gamma_1$ ($\Gamma_1/2\pi$ < 40 Hz), the frequencies of two mechanical resonators are distinguishable and the linewidths are almost degenerate (blue and black circles in Figs. 3a and 3b), with the frequencies of eigenmodes $\omega_\pm = (\omega_1+\omega_2)/2 \pm [(\delta\omega/2)^2 - \Gamma_1^2]^{1/2}$, where $\gamma_1 \approx \gamma_2 = \gamma$ is assumed. After the EP, two mechanical eigenmodes are frequency-degenerate, but separable in linewidth with $\gamma_\pm = \gamma +\Gamma_1 \mp [\Gamma_1^2 - (\delta\omega/2)^2]^{1/2}$, namely the dark mode (yellow circles) and the bright mode (red circles). The bright mode is measured with the cavity field instead of the probe beam for the better signal-to-noise ratio. The linewidth of the dark mode decreases continuously and gradually approaches the linewidth of the bare mode as $\Gamma_1$ increases, while the linewidth of the bright mode significantly increases, which is consistent with the experimental results obtained in the microwave optomechanical system [21]. After the second cavity mode $a_2$ is turned on, the mechanical dark mode is broken and the linewidth of the dark mode increases as $\Gamma_{12}$ increases, since the cavity mode $a_2$ provides an effective coupling between the dark and bright modes and the dark mode is cooled with the additional cavity mode. In the right panel of Figs. 3a and 3b, only the dark mode is plotted. The bright mode is continuously cooled by the second cavity mode. The downshift in frequency in Fig. 3a (purple circles) is due to the fact that the additional cavity mode destroys the mechanical degeneracy and splits the energy level, similar to the AC Stark shift in atomic systems.

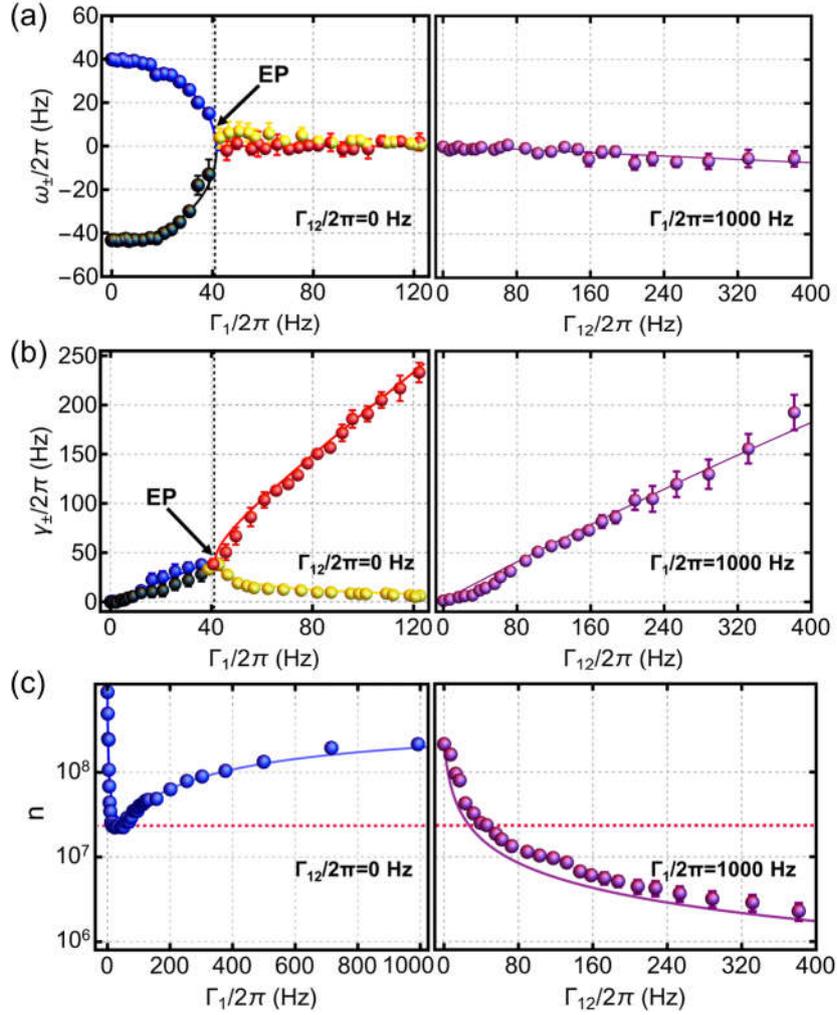

**Fig. 3.** (a) Frequency, (b) linewidth, and (c) total phonon number of the mechanical resonators as a function of $\Gamma_1$ and $\Gamma_{12}$. The circles and solid lines are the experimental results and theoretical simulations, respectively. The red dashed line in (c) is the cooling limit with the dark mode.

Figure 3c shows the measured phonon numbers as a function of $\Gamma_1$ and $\Gamma_{12}$. The phonon number of the mechanical resonator is $n_{1,2} = m\omega_{1,2}\langle x_{1,2}^2\rangle/\hbar$, where m is the effective mass and $x_{1,2}$ represents the displacement of the mechanical resonator. As shown in Fig. 3c, the total phonon number ($n=n_1+n_2$) of two resonators first decreases and then rises after passing the EP, which indicates that the cooling process is suppressed by the generation of mechanical dark mode. The total phonon number with a single cavity mode at a given $\Gamma_1$ can be expressed as [40]

$$n = \frac{2n_{th}\gamma\left[4(\gamma+\Gamma_1)^2 + \delta\omega^2\right]}{(\gamma+\Gamma_1)[4(\gamma+\Gamma_1)^2 + \delta\omega^2 - 4\Gamma_1^2]}, \quad (4)$$

where $n_{th}$ represents the average thermal phonon occupation of the mechanical resonator at the environmental temperature (the temperature is increased above the room temperature by applying white noise on the piezos for achieving better PSD signal-to-noise ratio). Under the conditions of $\Gamma_1 \gg \gamma$ and $\delta\omega^2 \gg \Gamma_1\gamma$, we can obtain n ≈ $2n_{th}\gamma(4\Gamma_1^2+\delta\omega^2)/\Gamma_1\delta\omega^2$. The EP at $\Gamma_1 = \delta\omega/2$ gives the dark mode cooling limit (red dashed line in Fig. 3c). By utilizing the second cavity mode, the dark mode cooling limit is broken, and the mechanical modes are cooled to a minimum phonon number that is more than an order of magnitude lower than the situation without the dark-mode-breaking. The further cooling is unobserved due to the limited detection sensitivity of the probe beam. The probe beam is necessary for observing the dark mode breaking process, which cannot be realized with a cavity field.

Although the optomechanical dark mode breaking cooling is achieved by choosing $g_{22} = 0$, it doesn't give the most efficient cooling process. By letting the optomechanical coupling vectors $\vec{G}_1 = (G_{11}, G_{21})$ and $\vec{G}_2 = (G_{21}, G_{22})$ to be perpendicular, the driving strength could be much smaller for reaching the same phonon number [29]. For this purpose, the single photon optomechanical coupling strengths are precisely tuned to satisfy the conditions $g_{11} = g_{21} \neq 0$, and $g_{12} = -g_{22} \neq 0$. By controlling the cavity input powers $P_1^{in}$ and $P_2^{in}$, we can make $\Gamma_1 + \Gamma_2 = 0$, and consequently, the non-diagonal terms in Eq. (2) are always zero and the dark mode is completely suppressed.

Figure 4a depicts the corresponding PSD of the cooling process of two membrane resonators. As $\Gamma_1$ and $\Gamma_2$ increase simultaneously with the fixed ratio to keep $\Gamma_1 + \Gamma_2 = 0$, both mechanical modes show a gradual decrease in amplitude and are not approaching each other. The frequencies and linewidths of the eigenmodes can be extracted, and they are plotted in Figs. 4b and 4c, respectively. In Fig. 4b, the frequencies of two mechanical modes remain unchanged because the effective interaction between mechanical modes induced by the two cavity modes is perfectly canceled. In Fig. 4c, the linewidths keep increasing, achieving the simultaneous cooling of two mechanical modes. $\delta\omega = 2\pi \times 60$ Hz is chosen for the purpose of clearly illuminating the phenomenon, and |δω| can be arbitrary small.

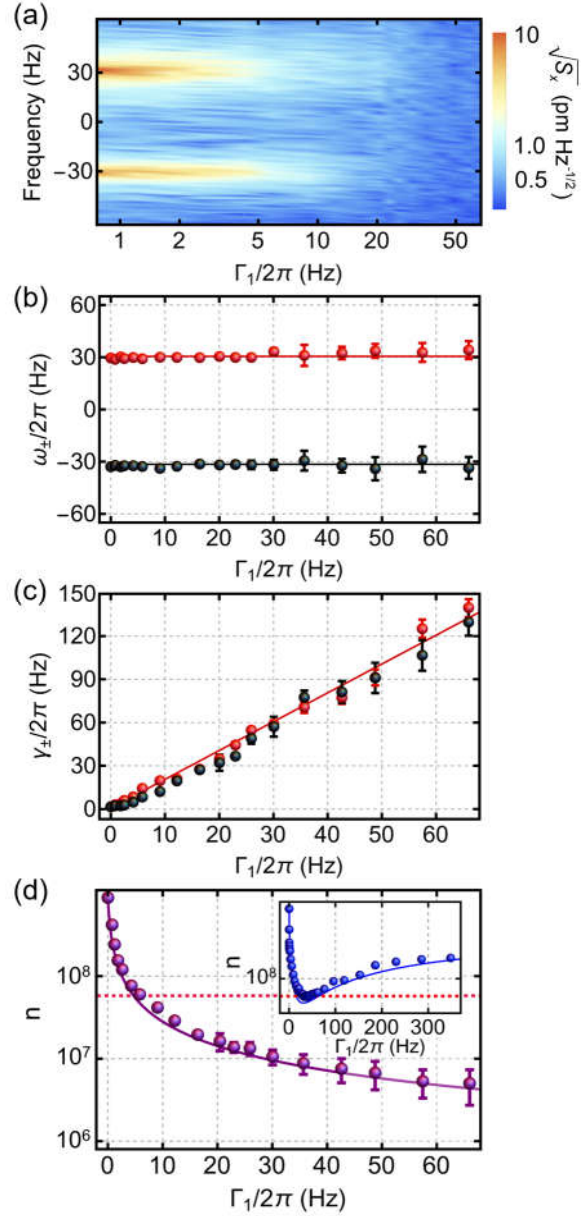

**Fig. 4.** (a) Measured PSD detected by the 1550 nm probe beam, (b) frequency, (c) linewidth, and (d) total phonon number of the mechanical resonators as a function of $\Gamma_1$. The circles and solid lines are the experimental results and theoretical simulations, respectively. The red dashed line in (d) is the cooling limit with the dark mode. The inset in (d) is the evolution of the phonon number when only one of the cavity modes is applied.

Compared to the single cavity mode cooling, two mechanical modes obviously experience a higher optomechanical damping rate, due to the fact that both two cavity modes act simultaneously on the mechanical modes with damping rates $\Gamma_{11}+\Gamma_{12}$ and $\Gamma_{21}+\Gamma_{22}$, respectively. Figure 4d shows the measured phonon number as a function of $\Gamma_1$. The phonon number keeps decreasing and is not limited by the dark mode cooling limit (red dashed line). The inset of Fig. 4d of the phonon number as a function of $\Gamma_1$ at only one of the cavity modes being applied displays the different dark mode cooling limit due to different frequency difference from Fig. 3c. The minimum achieved phonon number is more than one order of magnitude lower than the dark mode cooling limit, but at a much smaller driving strength compared to Fig. 3c.

In conclusion, we have presented a proof-of-principle experimental demonstration that the dark mode cooling limit can be broken by introducing a second cavity mode in a typical two-membrane cavity optomechanical system. It is worth mentioning that the second cavity mode can form another mechanical dark mode, and the dark modes generated by the two cavity modes can interfere destructively or constructively, which provides an effective way for the mechanical dark mode manipulation, not just for dark mode breaking. For example, by simply switching the sign of the optomechanical coupling coefficient to keep $\Gamma_1 = \Gamma_2$ instead of $\Gamma_1 = -\Gamma_2$, the dark mode effect will be enhanced rather than suppressed. Our experimental results can be straightforwardly generalized to the systems with N (N$\geq$2) degenerate mechanical modes by including N independent optical modes, thereby supporting the extension of quantum control to higher system dimensions. Our method is universal and could be generalized to other types of optomechanical systems, e.g. to control the vectorial polaritons in a levitated nanosphere [49]. Moreover, our system holds great promise for investigating macroscopic ground state cooling and quantum entanglement of multiple degenerate mechanical resonators, which may be essential for quantum gravity measurements in a multimode optomechanical system.

[1] M. Aspelmeyer, T. J. Kippenberg, and F. Marquardt, Cavity optomechanics, Rev. Mod. Phys. **86**, 1391 (2014).

[2] Jasper Chan, T. P. Mayer Alegre, Amir H. Safavi-Naeini, Jeff T. Hill, Alex Krause, Simon Gröblacher, Markus Aspelmeyer, and Oskar Painter, Laser cooling of a nanomechanical oscillator into its quantum ground state, Nature **478**, 89 (2011).

[3] Massimiliano Rossi, David Mason, Junxin Chen, Yeghishe Tsaturyan, and Albert Schliesser, Measurement-based quantum control of mechanical motion, Nature **563**, 53 (2018).

[4] T. P. Purdy, P.-L. Yu, R. W. Peterson, N. S. Kampel, and C. A. Regal, Strong optomechanical squeezing of light, Phys. Rev. X **3**, 031012 (2013).

[5] Amir H. Safavi-Naeini, Simon Gröblacher, Jeff T. Hill, Jasper Chan, Markus Aspelmeyer, and Oskar Painter, Squeezed light from a silicon micromechanical resonator, Nature **500**, 185 (2013).

[6] Nancy Aggarwal, Torrey J. Cullen, Jonathan Cripe, Garrett D. Cole, Robert Lanza, Adam Libson, David Follman, Paula Heu, Thomas Corbitt, and Nergis Mavalvala, Room-temperature optomechanical squeezing, Nat. Phys. **16**, 784 (2020).

[7] R. W. Andrews, R. W. Peterson, T. P. Purdy, K. Cicak, R. W. Simmonds, C. A. Regal, and K. W. Lehnert, Bidirectional and efficient conversion between microwave and optical light, Nat. Phys. **10**, 321 (2014).

[8] Moritz Forsch, Robert Stockill, Andreas Wallucks, Igor Marinković, Claus Gärtner, Richard A. Norte, Frank van Otten, Andrea Fiore, Kartik Srinivasan, and Simon Gröblacher, Microwave-to-optics conversion using a mechanical oscillator in its quantum ground state, Nat. Phys. **16**, 69 (2020).

[9] Prasoon K. Shandilya, David P. Lake, Matthew J. Mitchell, Denis D. Sukachev, and Paul E. Barclay, Optomechanical interface between telecom photons and spin quantum memory, Nat. Phys. **17**, 1420 (2021).

[10] Shlomi Kotler, Gabriel A. Peterson, Ezad Shojaee, Florent Lecocq, Katarina Cicak, Alex Kwiatkowski, Shawn Geller, Scott Glancy, Emanuel Knill, Raymond W. Simmonds, José Aumentado, and John D. Teufel, Direct observation of deterministic macroscopic entanglement, Science **372**, 622 (2021).

[11] L. M. de Lépinay, C. F. Ockeloen-Korppi, M. J. Woolley, and M. A. Sillanpää, Quantum mechanics-free subsystem with mechanical oscillators, Science **372**, 625 (2021).

[12] J. Piotrowski, D. Windey, J. Vijayan, C. Gonzalez-Ballestero, A. de los Ríos Sommer, N. Meyer, R. Quidant, O. Romero-Isart, R. Reimann, and L. Novotny, Simultaneous ground-state cooling of two mechanical modes of a levitated nanoparticle, Nat. Phys. **19**, 1009 (2023).

[13] S. Bose, A. Mazumdar, G. W. Morley, H. Ulbricht, M. Toroš, M. Paternostro, A. A. Geraci, P. F. Barker, M. S. Kim, and G. Milburn, Spin Entanglement Witness for Quantum Gravity, Phys. Rev. Lett. **119**, 240401 (2017).

[14] C. Marletto and V. Vedral, Gravitationally Induced Entanglement between Two Massive Particles is Sufficient Evidence of Quantum Effects in Gravity, Phys. Rev. Lett. **119**, 240402 (2017).

[15] Daniel Carney, Philip C E Stamp, and Jacob M Taylor, Tabletop experiments for quantum gravity: a user's manual, Class. Quantum Grav. **36**, 034001 (2019).

[16] Tobias Westphal, Hans Hepach, Jeremias Pfaff, and Markus Aspelmeyer, Measurement of gravitational coupling between millimetre-sized masses, Nature **591**, 225 (2021).

[17] J. Huang, D.-G. Lai, and J.-Q. Liao, Thermal-noise-resistant optomechanical entanglement via general dark-mode control, Phys. Rev. A **106**, 063506 (2022).

[18] D.-G. Lai, A. Miranowicz, and F. Nori, Noise-Tolerant Optomechanical Entanglement via Synthetic Magnetism, Phys. Rev. Lett. **129**, 063602 (2022).

[19] C. Genes, D. Vitali, and P. Tombesi, Simultaneous cooling and entanglement of mechanical modes of amicromirror in an optical cavity, New J. Phys. **10**, 095009 (2008).

[20] A. B. Shkarin, N. E. Flowers-Jacobs, S. W. Hoch, A. D. Kashkanova, C. Deutsch, J. Reichel, and J. G. E. Harris, Optically Mediated Hybridization between Two Mechanical Modes, Phys. Rev. Lett. **112**, 013602 (2014).

[21] C. F. Ockeloen-Korppi, M. F. Gely, E. Damskägg, M. Jenkins, G. A. Steele, and M. A. Sillanpää, Sideband cooling of nearly degenerate micromechanical oscillators in a multimode optomechanical system, Phys. Rev. A **99**, 023826 (2019).

[22] M. Fleischhauer, A. Imamoglu, and J. P. Marangos, Electromagnetically induced transparency: Optics in coherent media, Rev. Mod. Phys. **77**, 633 (2005).

[23] E. Arimondo, Coherent population trapping in laser spectroscopy, Prog. Opt. **35**, 257 (1996).

[24] K.-J. Boller, A. Imamoğlu, and S. E. Harris, Observation of electromagnetically induced transparency, Phys. Rev. Lett. **66**, 2593 (1991).

[25] Min Xiao, Yong-qing Li, Shao-zheng Jin, and Julio Gea-Banacloche, Measurement of Dispersive Properties of Electromagnetically Induced Transparency in Rubidium Atoms, Phys. Rev. Lett. **74**, 666 (1995).

[26] S. Weis, R. Rivière, S. Deléglise, E. Gavartin, O. Arcizet, A. Schliesser, and T. J. Kippenberg, Optomechanically Induced Transparency, Science **330**, 1520 (2010).

[27] J. D. Teufel, D. Li, M. S. Allman, K. Cicak, A. J. Sirois, J. D. Whittaker, and R. W. Simmonds, Circuit cavity electromechanics in the strong-coupling regime, Nature **471**, 204 (2011).

[28] C. Dong, V. Fiore, M. C. Kuzyk, and H. Wang, Optomechanical dark mode, Science **338**, 1609 (2012).

[29] J.-Y. Liu, W. Liu, D. Xu, J.-C. Shi, H. Xu, Q. Gong, and Y.-F. Xiao, Ground-state cooling of multiple near-degenerate mechanical modes, Phys. Rev. A **105**, 053518 (2022).

[30] J. Huang, D.-G. Lai, C. Liu, J.-F. Huang, F. Nori, and J.-Q. Liao, Multimode optomechanical cooling via general dark-mode control, Phys. Rev. A **106**, 013526 (2022)

[31] D.-G. Lai, J.-F. Huang, X.-L. Yin, B.-P. Hou, W. Li, D. Vitali, F. Nori, and J.-Q. Liao, Nonreciprocal ground-state cooling of multiple mechanical resonators, Phys. Rev. A **102**, 011502(R) (2020).

[32] M. T. Naseem and Ö. E. Müstecaplıoglu, Ground-state cooling of mechanical resonators by quantum reservoir engineering, Commun. Phys. **4**, 95 (2021).

[33] C. Sommer and C. Genes, Partial Optomechanical Refrigeration via Multimode Cold-Damping Feedback, Phys. Rev. Lett. **123**, 203605 (2019).

[34] Z. Yang, J. Yang, S.-L. Chao, C. Zhao, R. Peng, L. Zhou, Simultaneous ground-state cooling of identical mechanical oscillators by Lyapunov control, Opt. Express **30,** 20145 (2022).

[35] Pengyu Wen, Xuan Mao, Min Wang, Chuan Wang, Gui-Qin Li, and Gui-Lu Long, Simultaneous ground-state cooling of multiple degenerate mechanical modes through the cross-Kerr effect, Opt. Lett. **47**, 5529 (2022).

[36] X. Wei, J. Sheng, C. Yang, Y. Wu, and H. Wu, Controllable two-membrane-in-the-middle cavity optomechanical system, Phys. Rev. A **99**, 023851 (2019).

[37] J. Sheng, X. Wei, C. Yang, and H. Wu, Self-Organized Synchronization of Phonon Lasers, Phys. Rev. Lett **124**, 053604 (2020).

[38] P. Piergentili, L. Catalini, M. Bawaj, S. Zippilli, N. Malossi, R. Natali, D. Vitali, and G. D. Giuseppe, Two-membrane cavity optomechanics, New J. Phys. **20**, 083024 (2018)

[39] Q. Zhang, C. Yang, J. Sheng, and H. Wu, Dissipative coupling-induced phonon lasing, PNAS **119**, 52 (2022).

[40] See Supplemental Material for additional information about the theory, numerical simulations, and experimental details, which includes Refs. [1, 3, 28-30, 37-38, 41-47].

[41] C. Zhu and B. Stiller, Dynamic Brillouin cooling for continuous optomechanical systems, Mater. Quantum Technol. 3, 015003 (2023).

[42] Shuhui Wu, Jiteng Sheng, Xiaotian Zhang, Yuelong Wu, Haibin Wu, Parametric excitation of a SiN membrane via piezoelectricity, AIP Advances 8, 015209 (2018).

[43] M. Underwood, D. Mason, D. Lee, H. Xu, L. Jiang, A. B. Shkarin, K. Børkje, S. M. Girvin, and J. G. E. Harris, Measurement of the motional sidebands of a nanogram-scale oscillator in the quantum regime, Phys. Rev. A 92, 061801(R) (2015).

[44] T. P. Purdy, P.-L. Yu, N. S. Kampel, R. W. Peterson, K. Cicak, R. W. Simmonds, and C. A. Regal, Optomechanical Raman-ratio thermometry, Phys. Rev. A 92, 031802(R) (2015).

[45] J. Li, A. Xuereb, N. Malossi, and D. Vitali, Cavity mode frequencies and strong optomechanical coupling in two-membrane cavity optomechanics, J. Opt. 18, 084001 (2018).

[46] V. Fedoseev, F. Luna, I. Hedgepeth, W. Löffler, and D. Bouwmeester, Stimulated Raman Adiabatic Passage in Opotomechanics, Phys. Rev. Lett. **126**, 113601 (2021).

[47] Cheng Yang, Xinrui Wei, Jiteng Sheng, and Haibin Wu, Phonon heat transport in cavity-mediated optomechanical nanoresonators. Nat. Commun. **11**, 4656 (2020).

[48] M. A. Miri and A. Alù, Exceptional points in optics and photonics, Science **363**, 7709 (2019).

[49] A. Ranfagni, P. Vezio, M. Calamai, A. Chowdhury, F. Marino, and F. Marin, Vectorial polaritons in the quantum motion of a levitated nanosphere, Nature Physics **17**, 1120-1124 (2021).